\documentclass[12pt]{article}
\usepackage{amsmath}
\usepackage{graphicx,psfrag,epsf}
\usepackage{enumerate}
\usepackage{natbib}
\usepackage{amssymb}
\usepackage{nccmath}
\usepackage{xcolor}
\usepackage{multirow}
\usepackage{booktabs}
\usepackage{mathtools}
\usepackage{float}
\usepackage{appendix}

\newtheorem{theorem}{Theorem}

\allowdisplaybreaks

\newcommand{\ind}{\perp\!\!\!\!\perp}
\newcommand{\blind}{0}

\usepackage{titlesec}
\titleformat*{\section}{%
    \fontsize{12}{12}\bfseries%
}

\titleformat*{\subsection}{%
    \fontsize{11}{11}\bfseries%
}

\begin{document}

\def\spacingset#1{\renewcommand{\baselinestretch}%
{#1}\small\normalsize} \spacingset{1}

\if0\blind
{
  \title{\bf A New Targeted-Federated Learning Framework for Estimating Heterogeneity of Treatment Effects: A Robust Framework with Applications in Aging Cohorts}
  \author{Rong Zhao$^{1}$, Jason Falvey$^{2}$, Xu Shi$^{3}$,
Vernon M. Chinchilli$^{1,\dagger}$, and
Chixiang Chen$^{4,\dagger}$\\
$^{1}$Department of Public Health Sciences, Penn State College of Medicine\\
$^{2}$Department of Physical Therapy and Rehabilitation Science, \\University of Maryland School of Medicine\\
$^{3}$Department of Biostatistics, University of Michigan, Ann Arbor\\
$^{4}$Department of Epidemiology and Public Health,\\ University of Maryland School of Medicine\\
$^{\dagger}$ Co-corresponding authors}
  \maketitle
} \fi

\if1\blind
{
  \bigskip
  \bigskip
  \bigskip
  \begin{center}
    {\LARGE\bf Title}
\end{center}
  \medskip
} \fi

\bigskip
\begin{abstract}
{Analyzing data from multiple sources offers valuable opportunities to improve the estimation efficiency of causal estimands for a target population. However, this analysis also poses many challenges due to population heterogeneity and data privacy constraints. While several advanced methods for causal inference in federated settings have been developed in recent years, most focus on average causal effects based on mean differences} and are not designed to study effect modification for a target population. In this paper, we introduce a novel targeted-federated learning framework to study the heterogeneity of treatment effects (HTEs) for a targeted population through a working structural model. This framework integrates information from multiple data sources without sharing raw data, while accounting for covariate distribution shifts from the target population. Our proposed approach is shown to be doubly robust, conveniently supporting the interpretation of distinct outcome types. Furthermore, we develop a communication-efficient bootstrap-based selection algorithm to detect non-transportable data sources, thereby enhancing robust information aggregation. Through extensive simulation studies, we demonstrate the superior performance of the proposed estimator, in terms of estimation bias and efficiency gain relative to other methods. We further illustrate the practical utility of our approach in a real-world data application using nationwide Medicare-linked data from an aging cohort.
\end{abstract}

\noindent%

\spacingset{1.9}
\section{INTRODUCTION}

Personalized medicine aims to tailor healthcare decisions to individual patient characteristics, recognizing that treatment effects often vary across different subpopulations. This variation, known as heterogeneity of treatment effects (HTEs), has critical implications for clinical practice, health policy, and equitable care delivery. A one-size-fits-all estimate of treatment efficacy, such as the average treatment effect (ATE), may mask important differences in how subgroups respond to interventions, potentially leading to suboptimal or even harmful decisions for certain individuals. As a result, there is growing emphasis on identifying and quantifying HTEs to better inform personalized treatment strategies.

From a statistical perspective, HTEs can be formulated as treatment effects that vary as a function of covariates \citep{kennedy2023towards, morzywolek2024weightedorthogonallearnersheterogeneous, nie2021quasi}. In the past decade, an increasing number of methods have been developed to better characterize these effects, including the T-learner \citep{kunzel2019metalearners}, R-learner \citep{nie2021quasi}, and DR-learner \citep{kennedy2023towards}, among others. Most of these approaches primarily focus on difference-based estimands for continuous outcomes, although with extensions to other outcome types, such as conditional estimands for time-to-event outcomes, have also been studied \citep{morzywolek2024weightedorthogonallearnersheterogeneous}. In observational studies, estimating HTEs typically requires adjustment for confounding, usually through modeling the propensity score (PS), the outcome regression (OR) model, or both. Effect modification is then characterized separately through an additional model for the underlying HTE as a function of certain covariates \citep{nie2021quasi, kennedy2023towards}. 

Furthermore, most existing literature on HTEs relies on a single dataset, which often limits statistical power due to small sample sizes, particularly in a single-site study. With the growing availability of clinical data across institutions, there is a burgeoning interest in leveraging multi-source data to enhance the estimation of HTEs. When data are accessible across sources, data fusion approaches have been developed to enable semiparametric modeling for combining information across multiple sources \citep{li2023efficient, li2024efficient, li2025data, yang2025data}. 
When privacy regulations or institutional constraints prohibit such sharing, federated learning offers a compelling alternative, enabling collaborative analysis across multiple institutions/study groups/study years without the need to pool individual-level data \citep{dayan2021federated}. By preserving data privacy and complying with institutional or regulatory constraints, federated learning facilitates large-scale inference while mitigating the risks and complexities associated with centralized data sharing. While the majority of federated learning methods to date have focused on prediction \citep{brisimi2018federated, duan2022heterogeneity, li2023targeting, gu2023commute} and classification tasks \citep{hsieh2014divide, wang2019distributed}, recent advancements have extended federated learning frameworks to support causal inference \citep{vo2021federated, xiong2023federated,hu2024collaborative,liu2025federated}. However, these approaches primarily estimate the ATE for the overall study population, rather than focusing on a prespecified target population that is often of greatest interest to researchers. Such target populations may include patients admitted in the most recent year, subpopulations from specific geographic regions, individuals with particular baseline characteristics or comorbidities, or historically underrepresented groups in clinical research \citep{lesko2017generalizing, han2024privacy}. Tailoring causal inference to these specific populations can yield more relevant and actionable insights for decision-making and policy development. More recently, \citep{han2024privacy,han2025federated} developed a targeted-federated learning approach for the mean potential outcome. They further proposed a transfer learning method based on influence functions of various estimators to facilitate robust data integration. To the best of our knowledge, however, a significant methodological gap remains in the literature regarding the development of targeted-federated learning approaches for robustly estimating HTEs, beyond the averaged effect.

To address these methodological gaps, we develop a novel targeted-federated learning framework for estimating HTEs within a prespecified targeted population. We formally define the underlying HTE using a projection-based estimand, which is broadly applicable across outcome types, including count and binary outcomes. This estimand builds on a working structural model that can serve as a reasonable approximation to the underlying HTE function. Building on this framework, we develop a doubly robust estimator that integrates information from multiple datasets under the targeted-federated paradigm. The estimator is doubly robust in the sense that it yields a consistent estimate if either the OR model is correctly specified or both the PS and density ratio models are correctly specified, but not necessarily all three. The density ratio model plays a crucial role in calibrating covariate distribution shifts across datasets \citep{qin1998inferences, han2024privacy, chen2024integrating}. To further safeguard against negative transfer, we introduce a communication-efficient bootstrap-based selection procedure designed to detect and exclude non-transportable data sources that could bias estimation. This procedure aims to incorporate only compatible information and empirically improves the robustness of targeted-federated learning. 

The proposed framework has two key innovations that distinguish it from existing literature. First, integrating the proposed working structural model with the targeted-federated learning framework yields several methodological advancements: the resulting estimator is doubly robust, accommodates covariate shifts across datasets, and requires only a single round of information exchange, making it both communication-efficient and easy to implement. It also provides interpretable estimates of effect modification and enables computationally feasible estimation in federated learning settings without sharing raw data. Second, the proposed dataset selection algorithm avoids the demanding calculations of influence functions under covariate shift and requires no additional communication rounds, thereby facilitating implementation in federated settings.

The remainder of the article is organized as follows. Section \ref{METHODOLOGY} introduces notation and the proposed targeted-federated estimators. Section \ref{SIMULATION} summarizes extensive simulation studies to demonstrate the finite-sample performance of the proposed methods. Section \ref{REAL DATA APPLICATION} describes the application of our proposed methods to nationwide Medicare claims data on older adults who experienced a hip fracture. Section \ref{DISCUSSION} discusses future directions and concludes the paper.

\section{METHODOLOGY}
\label{METHODOLOGY}
\subsection{Basic setup}
\label{Basic setup}
Consider data on $N$ individuals stored across $K+1$ independent data sources, indexed by the set $\mathcal{K}=\{1, \dots, K+1\}$. The $m$-th data source has a sample size of $n_m$, such that $N=\sum_{m=1}^{K+1} n_m$. Let $M_i=m$ be the source indicator for subject $i$, indicating that their data are stored at source $m\in \mathcal{K}$. For notational convenience, we designate the data from the first source ($M=1$) as the target dataset of interest and treat the remaining sources (i.e., $m\in\mathcal{S}=\{2,\dots,K+1\}$) as external datasets. For each subject $i$, we observe an outcome of interest $Y_i$, which may be on either a continuous or binary scale (e.g., stay more days at home (DAH) vs fewer), a binary treatment indicator $A_i\in\{0, 1\}$ (e.g., surgical types: open reduction with internal fixation (OR/IF) vs total hip arthroplasty (HA)), and a $p$-dimensional vector of baseline covariates $\mathbf{X}_i=(X_{i1},\dots,X_{ip})^\top$ (e.g., age, sex, comorbid conditions, etc.). The distributions of baseline covariates may vary across datasets,  with source-specific density functions denoted by $f(\mathbf{X}|M=m)$ for $m\in \mathcal{K}$. The observed data from the $m$-th source are summarized as $D_m=\big\{(Y_i, \mathbf{X}_i^\top, A_i)^\top: 1\leq i \leq N ~\text{and}~ M_i=m \big\}$. 

This paper addresses a practical scenario in which sharing individual-level data across sources is challenging. In the case study, we aim to evaluate the differential impact of two surgical procedures for older hip fracture survivors. Our research team has access to Medicare fee-for-service data from 2017, which serves as the target dataset. To strengthen statistical power, we collaborate with a partner who has additional historical data from 2010 to 2016. This collaborative framework naturally motivates the use of targeted-federated learning. 

Before concluding this section, we describe the Neyman-Rubin causal paradigm \citep{splawa1990application,rubin2005causal}. Let $Y_i(a)$ be the potential outcome for individual $i$ under treatment assignment $A_i = a$, with $a \in \{0, 1\}$. Only one of the two potential outcomes is observed for each individual. The observed outcome is linked to the potential outcomes through $Y_i=A_iY_i(1)+(1-A_i)Y_i(0)$. To enable valid causal inference, we adopt the following standard assumptions, collectively known as the Stable Unit Treatment Value Assumption (SUTVA):

\textbf{Assumption 1} (No interference) The potential outcomes of any individual are unrelated to the treatment assignment of other individuals.

\textbf{Assumption 2} (Consistency) For each individual $i$, if $A_i=a$, $Y_i=Y_i(a)$.

Assumption 1 requires no interference between individuals, which is reasonable in our case study since most of the Medicare beneficiaries in our dataset who experience a hip fracture are treated and live independently. Assumption 2 states that the observed outcome is equal to the potential outcome under the assigned treatment, and there are no multiple versions of treatment. This is plausible in our application, as OR/IF and HA are clinically distinct, well-defined procedures, each uniquely identified by specific procedure codes in Medicare claims data, leaving no ambiguity in the binary treatment indicator.

\vspace{-1.5ex}

\subsection{Heterogeneity of treatment effects via a working structural model}\label{Heterogeneity of Treatment Effect via A Working Structural Model}

One of the primary goals of HTE analysis is to assess how treatment effects vary according to certain baseline covariates (e.g., sex and chronic conditions, $\mathbf{\tilde{X}} \subset \mathbf{X}$). To enhance applicability and promote more interpretable results analogous to conventional regression, 
we define our target estimand $\boldsymbol{\beta}_0$ for HTE analysis as the parameter in the structural model $l^{-1}\{\boldsymbol{\eta}(\mathbf{\mathbf{\tilde{X}}},a)^\top\boldsymbol{\beta}_0\}$ solving
\begin{equation}\label{beta0 def}
    E\bigg(\sum_{a\in\{0,1\}}\boldsymbol{\eta}(\mathbf{\tilde{X}},a)\Big[Y(a)-l^{-1}\{\boldsymbol{\eta}(\mathbf{\tilde{X}},a)^\top\boldsymbol{\beta}_0\}\Big]\Big\lvert M=1\bigg)=\mathbf{0},
\end{equation}
where $l(\cdot)$ is a link function, which could be a log-link if $Y(a)$ is count data, a logit-link if $Y(a)$ is binary, or an identity link if $Y(a)$ is continuous. $\boldsymbol{\eta}(\mathbf{\tilde{X}},a)$ denotes a pre-specified vector involving $\mathbf{\tilde{X}}$ and $a$, with the same dimension as $\boldsymbol\beta_0$. The model is structural in that it directly models the potential outcomes. Below, we provide motivation and interpretation for the proposed estimand.

In many comparative effectiveness studies, effect modifiers are often categorical and the number of candidate variables under consideration is typically small. Under such settings, the structural model defined in (\ref{beta0 def}) may coincide with the conditional expectation, i.e., $l\{E(Y(a)|\mathbf{\tilde{X}}, M=1)\}=\boldsymbol{\eta}(\mathbf{\mathbf{\tilde{X}}},a)^\top\boldsymbol{\beta}_0$. For example, with a binary effect modifier $X_1$ (e.g., sex), taking the working form of $\boldsymbol{\eta}(\mathbf{\tilde{X}},a)=(1,a,X_1,aX_1)^\top$ and a logit link, the model is saturated, and the components of $\boldsymbol{\beta}_0$ excluding the intercept represent the log odds ratios for treatment, sex and effect modification by sex. 
When we set $\boldsymbol{\eta}(\mathbf{\tilde{X}},a)^\top \boldsymbol{\beta}_0=\beta_1+\beta_2 a$, it reduces to the commonly used parametric marginal structural models (MSMs) \citep{robins2000marginal} for the average treatment effect estimand.

When $\mathbf{\tilde{X}}$ contains several continuous variables, however, the conditional surface $l\{E(Y(a)|\mathbf{\tilde{X}}, M=1)\}$ may not precisely equal the structural model $\boldsymbol{\eta}(\mathbf{\tilde{X}},a)^\top\boldsymbol{\beta}_0$. This is well documented in the HTE literature, where the underlying truth cannot always be explicitly captured \citep{semenova2021debiased, morzywolek2024weightedorthogonallearnersheterogeneous}. 
In such settings, the proposed working structural model in (\ref{beta0 def}), which aims to reasonably approximate the potential outcomes, serves as a practical alternative.
Intuitively, the working function $\boldsymbol{\eta}(\mathbf{\tilde{X}},a)^\top\boldsymbol{\beta}_0$ can be interpreted as the generalized linear projection of the potential outcome $Y(a)$ onto the space spanned by $\{\mathbf{\tilde{X}},a\}$, analogous to a conventional generalized linear regression as if $Y(a)$ for $a\in\{0,1\}$ were observed. 
In the case of a continuous outcome with the identity link, for example, the moment condition \eqref{beta0 def} is then the first-order condition of minimizing population-level squared error loss. Therefore, $\boldsymbol{\beta}_0$ can be interpreted as the best linear predictor of $Y(a)$ within the target population in the squared-error sense \citep{semenova2021debiased}, analogous to ordinary least squares under model misspecification. Further discussion of interpretation under other links and outcomes is provided in Section~1.1 of the Supplementary Material.

\subsection{Identification and estimation}\label{The Proposed Estimation}

In this section, we introduce the identification of the estimand and a robust estimator based solely on the target data, followed by an estimation procedure that integrates data from multiple sources in a federated setting. Then, we establish the theoretical properties of the proposed methods and discuss an extension to scenarios in which data from certain sources should not be borrowed.

\subsubsection{Identification and doubly robust estimator using the target data}\label{Doubly Robust Estimator Using Target Data}

To identify the proposed estimand, one could use the target source data alone based on the working structural model described in the previous section. To proceed, we make two additional assumptions. The discussions of these assumptions can be found in Section \ref{REAL DATA APPLICATION} and the Supplementary Material Section 3.1.

\textbf{Assumption 3} (Positivity of treatment assignment in the target site): $0<Pr(A=1|\mathbf{X}=\mathbf{x}, M=1)<1$, for almost every $\mathbf{x}$ such that $f(\mathbf{x}|M=1)>0$.

\textbf{Assumption 4} (Ignorability in the target site): $Y(a)\ind A|\mathbf{X},M=1$ for $a\in\{0,1\}$ and all individuals in the target site.

Under Assumptions 1 to 4, it is straightforward to identify the underlying target $\boldsymbol{\beta}_0$ by solving the estimating equations using the observed data
    \begin{equation*}
        E\!\left(\sum_{a\in\{0,1\}} \boldsymbol{\eta}(\tilde{\mathbf{X}},a)\Big[E(Y|A=a,\mathbf{X},M=1) - l^{-1}\!\left\{\boldsymbol{\eta}(\tilde{\mathbf{X}},a)^\top\boldsymbol{\beta}_0\right\}\Big]\,\bigg|\,M=1\right) = \mathbf{0}.
    \end{equation*} The underlying target $\boldsymbol{\beta}_0$ can be estimated by the doubly robust estimator $\hat{\boldsymbol{\beta}}_T$, obtained by solving the following estimating equations:

\begin{equation}
\label{Eq:dr_est_target}
\begin{split}
    n_1^{-1}\sum_{i=1}^N I(M_i=1)\Biggl(\sum_{a\in \{0,1\}}\boldsymbol{\eta}(\boldsymbol{\mathbf{\tilde{X}}}_i,a)\biggl[ & \frac{I_{\{A_i=a\}}}{\pi_1(a,\mathbf{X}_i;\hat{\boldsymbol{\gamma}}_1)}  
    \Bigl\{Y_i-g_1(a,\mathbf{X}_i;\hat{\boldsymbol{\psi}}_1)\Bigr\} \\ &+  \Bigl\{g_1(a,\mathbf{X}_i;\hat{\boldsymbol{\psi}}_1)-l^{-1}\{\boldsymbol{\eta}^\top(\boldsymbol{\mathbf{\tilde{X}}}_i,a)\boldsymbol{\beta}_T\} \Bigr\} \biggr]\Biggl) = \boldsymbol{0}
\end{split}
\end{equation}
where $\pi_1(a,\mathbf{X}_i;\hat{\boldsymbol{\gamma}}_1)$ and $g_1(a, \mathbf{X}_i; \hat{\boldsymbol{\psi}}_1)$ are the estimated propensity score (PS) model and outcome regression (OR) model, respectively, with parameters $\hat{\boldsymbol{\gamma}}_1$ and $\hat{\boldsymbol{\psi}}_1$ estimated by maximum likelihood approaches. The functions $l(\cdot)$ and $\boldsymbol{\eta}(\cdot)$ have been defined in Section \ref{Heterogeneity of Treatment Effect via A Working Structural Model}. Equation~(\ref{Eq:dr_est_target}) can be viewed as a two-step doubly robust estimating equation. In the first step, the nuisance functions $\pi_1(a,\mathbf{X};\hat{\boldsymbol{\gamma}}_1)$ and $g_1(a,\mathbf{X};\hat{\boldsymbol{\psi}}_1)$ are fitted for confounding adjustment using the full covariate vector $\mathbf{X}$. In the second step, these fitted nuisance functions form an estimating equation that projects the resulting pseudo-outcome structure onto the working structural model indexed by $\boldsymbol{\eta}(\tilde{\mathbf{X}},a)$. The first term in Equation~(\ref{Eq:dr_est_target}) is an inverse-probability-of-treatment-weighted (IPTW) augmentation term, while the second term corresponds to conditional mean imputation (CMI) estimating equation, which imputes the unobserved conditional means at both $a=0$ and $a=1$. The estimator $\hat{\boldsymbol{\beta}}_T$ has two key features: first, the estimator $\hat{\boldsymbol{\beta}}_T$ possesses the doubly robust property, as formalized in Theorem \ref{doubly_robust_target}.

\begin{theorem}\label{doubly_robust_target}
Given Assumptions 1 to 4 and regularity conditions \citep{van2000asymptotic}, the estimator $\hat{\boldsymbol{\beta}}_T$ is consistent if either $E(Y|A=a, \mathbf{X}, M=1)=g_1(a, \mathbf{X}_i; \boldsymbol{\psi}_1^\ast)$ or $Pr(A=a|\mathbf{X}, M=1)=\pi_1(a,\mathbf{X}_i;\boldsymbol{\gamma}^\ast_1)$ holds for $a\in\{0,1\}$, but not necessarily both, with $\hat{\boldsymbol{\psi}}_1-\boldsymbol{\psi}_1^\ast=o_p(1)$ and $\hat{\boldsymbol{\gamma}}_1-\boldsymbol{\gamma}^*_1=o_p(1)$. Moreover, the estimator
 follows an asymptotically normal distribution when nuisance functions are estimated parametrically.
\end{theorem}

Second, the estimator $\hat{\boldsymbol{\beta}}_T$ does not have an analytical form in general, necessitating an iterative estimation scheme. However, the term $\bigl\{\boldsymbol{\eta}(\boldsymbol{\mathbf{\tilde{X}}}_i,a){I_{\{A_i=a\}}}/{\pi_1(a,\mathbf{X}_i;\hat{\boldsymbol{\gamma}}_1)}\bigr\} \bigl\{Y_i-g_1(a,\mathbf{X}_i;\hat{\boldsymbol{\psi}}_1)\bigr\}$ does not directly involve the primary parameters $\boldsymbol{\beta}_T$ and therefore does not need to be updated in each iteration. This feature is essential for devising an efficient distributed algorithm for our targeted-federated estimator, as described in the next section.

\subsubsection{Doubly robust targeted-federated learning}
\label{Doubly Robust Targeted-Federated Learning}
The strategy of using only target data may be underpowered to capture the heterogeneity of treatment effects due to the limited sample size and insufficient population diversity in a single data source. Integrating information from other data sources can help achieve a more precise estimate. We propose a doubly robust targeted-federated estimator that achieves three goals: (1) flexibly accommodating heterogeneity in covariate distributions across data sources, (2) preserving data privacy by eliminating the need to share individual-level data, and (3) minimizing the communication costs and the time required to transmit information across sites. To achieve these goals and promote non-parametric identification, we require additional assumptions for the source datasets: 

\textbf{Assumption 5} (Positivity of treatment assignment in source datasets): $0<Pr(A=1|\mathbf{X}=\mathbf{x}, M=m)<1$, for each $m\in \mathcal{S}$ and almost every $\mathbf{x}$ such that $f(\mathbf{x}|M=m)>0$.

\textbf{Assumption 6} (Ignorability in source datasets): $Y(a)\ind A|\mathbf{X},M=m$ for $a\in\{0,1\}$ and all individuals in source datasets $m\in \mathcal{S}$.

\textbf{Assumption 7} (Positive covariate distributions): For each $m\in\mathcal{S}$, $f(\mathbf{x}|M=m)>0$ for almost every $\mathbf{x}$ such that $f(\mathbf{x}|M=1)>0$.

\textbf{Assumption 8} (Mean exchangeability over data sources): $E(Y|A=a, \mathbf{X}=\mathbf{x})=E(Y|A=a, \mathbf{X}=\mathbf{x},M=m)$ for $m\in \mathcal{K}$, $a\in\{0,1\}$, and almost every $\mathbf{x}$ such that $f(\mathbf{x}|M=1)>0$.

Assumptions 5 and 6 extend Assumptions 3 and 4 to the source datasets. Assumption 7 additionally requires that the covariate distributions across different sites remain non-zero when $\mathbf{x}$ assumes possible values from the target data, which is essential for validating the use of the density ratio model (\cite{qin1998inferences}). Assumption 8 requires that the conditional mean outcome models are independent of the data source allocation. Further discussion of Assumptions 6 and 8 can be found in Section~1.5 of the Supplementary Material. 

We now present the estimation procedure. The proposed targeted-federated estimator, denoted by $\hat{\boldsymbol{\beta}}_F$, can be obtained by solving the following estimating equations:
\begin{equation}
\label{Eq:dr_targeted_federated}
    \sum_{m\in \mathcal{S}} w_m\mathbf{P}_m + w_1\mathbf{P}_1 + \mathbf{Q}_1 = \mathbf{0}, ~ \text{where}
\end{equation}
\begin{align*}
    \mathbf{P}_m & = \sum_{i=1}^N \sum_{a\in \{0,1\}}\frac{I(M_i=m)}{n_m}\boldsymbol{\eta}(\boldsymbol{\mathbf{\tilde{X}}}_i,a)\biggl[ \rho_m(\mathbf{X}_i;\hat{\boldsymbol{\alpha}}_m)\frac{I_{\{A_i=a\}}}{\pi_m(a,\mathbf{X}_i;\hat{\boldsymbol{\gamma}}_m)} \ 
    \Bigl\{Y_i-g_m(a,\mathbf{X}_i;\hat{\boldsymbol{\psi}}_m)\Bigr\} \biggr], \\
    \mathbf{P}_1 & = \sum_{i=1}^N \sum_{a\in \{0,1\}}\frac{I(M_i=1)}{n_1}\boldsymbol{\eta}(\boldsymbol{\mathbf{\tilde{X}}}_i,a)\biggl[\frac{I_{\{A_i=a\}}}{\pi_1(a,\mathbf{X}_i;\hat{\boldsymbol{\gamma}}_1)} \ 
    \Bigl\{Y_i-g_1(a,\mathbf{X}_i;\hat{\boldsymbol{\psi}}_1)\Bigr\} \biggr],\\
    \mathbf{Q}_1 & = \sum_{i=1}^N \sum_{a\in \{0,1\}}\frac{I(M_i=1)}{n_1}\boldsymbol{\eta}(\boldsymbol{\mathbf{\tilde{X}}}_i,a)\biggl[ g_1(a,\mathbf{X}_i;\hat{\boldsymbol{\psi}}_1)-l^{-1}\{\boldsymbol{\eta}^\top(\boldsymbol{\mathbf{\tilde{X}}}_i,a)\boldsymbol{\beta}_F \} \biggr].
\end{align*}
Here, $\pi_m(a,\mathbf{X}_i;\hat{\boldsymbol{\gamma}}_m)$ and $g_m(a, \mathbf{X}_i; \hat{\boldsymbol{\psi}}_m)$ are the estimated PS model and OR model, respectively, with parameters $\hat{\boldsymbol{\gamma}}_m$ and $\hat{\boldsymbol{\psi}}_m$ estimated using source dataset $m\in\mathcal{S}$. The proposed estimator specifies the dataset-specific weights $w_m=n_m/N$, for $m\in \mathcal{K}$ and aggregates the information from external source data by their sample sizes. Other non-negative weights $w_m$, satisfying $\sum_{m\in \mathcal{K}} w_m=1$ can also be selected. For example, when $w_1=1$ and $w_m=0$ for all $m\in \mathcal{S}$, this targeted-federated estimator becomes the target-only estimator described in Section \ref{Doubly Robust Estimator Using Target Data}. 

We now provide the rationale behind the proposed estimating equations, with a focus on addressing covariate distribution shifts, achieving the doubly robust property,  and ensuring efficient communication and computation. Specifically, to account for distributional shifts in covariates across datasets, we adopt the density tilting approach by incorporating a tilting function $\rho_m(\mathbf{X}_i;\boldsymbol{\alpha}_m)$, parameterized by $\boldsymbol{\alpha}_m$, which reweights the source populations ($m\in\mathcal{S}$) to align with the target population and has been widely used in recent literature \citep{qin1998inferences,duan2020fast,han2024privacy}. The detailed estimation procedure is described in Section 1.4 of the Supplementary Material. Moreover, the resulting estimator given by the proposed estimating equations preserves the doubly robust property as stated in the following theorem.

\begin{theorem}\label{doubly_robust_federated}
Given Assumptions 1 to 8 and regularity conditions \citep{van2000asymptotic}, the estimator $\hat{\boldsymbol{\beta}}_F$ is a consistent estimator if either (i) $E(Y|A=a, \mathbf{X}, M=m)=g_m(a, \mathbf{X}_i; \boldsymbol{\psi}^\ast_m)$, $m\in\mathcal{K}$ or (ii) $Pr(A=a|\mathbf{X}, M=m)=\pi_m(a,\mathbf{X}_i;\boldsymbol{\gamma}^\ast_m)$, $m\in\mathcal{K}$ and $f(\mathbf{X}|M=1)/f(\mathbf{X}|M=m)=\rho_m(\mathbf{X}_i;\boldsymbol{\alpha}^\ast_m)$, $m\in\mathcal{S}$ hold for $a\in\{0,1\}$, with $\hat{\boldsymbol{\psi}}_m-\boldsymbol{\psi}_m^\ast=o_p(1)$, $\hat{\boldsymbol{\gamma}}_m-\boldsymbol{\gamma}^*_m=o_p(1)$ and $\hat{\boldsymbol{\alpha}}_m-\boldsymbol{\alpha}_m^\ast=o_p(1)$. Moreover, the estimator follows an asymptotically normal distribution when nuisance functions are estimated parametrically.
\end{theorem}

Theorem \ref{doubly_robust_federated} holds not only when all models in $m\in\mathcal{K}$ satisfy condition (i) or all models in $m\in\mathcal{S}$ satisfy condition (ii), but also under broader modeling scenarios. We refer readers to Section~1.3 \& 1.5 of the Supplementary Material for more details. Additionally, the proposed method is communication-efficient, requiring only a single round of interaction without sharing individual-level data across sources. Notably, covariate summary statistics from the target site, such as the covariate means, are first shared with the source sites to estimate ${\boldsymbol{\alpha}}_m$ in the density tilting model, after which each source transfers its partial information $\mathbf{P}_m$ back to the target analysis to estimate ${\boldsymbol{\beta}}_F$. This single-round communication framework reduces the risk of analysis errors and minimizes the logistical burden of data integration.

\subsection{Method extension}\label{method extension}

The federated estimation framework introduced in Section \ref{Doubly Robust Targeted-Federated Learning} relies on several critical assumptions, including doubly robust conditions, i.e., that either the OR models or both the density ratio and PS models be correctly specified, and several assumptions, particularly for the assumption of mean exchangeability. Notably, substantial heterogeneity often persists across source-specific populations, clinical practices, or healthcare policies. Integrating information from such non-transportable sources may bias estimation, a phenomenon often referred to as ``negative transfer" in the machine learning literature \citep{weiss2016survey}. Here, non-transportable datasets are defined as datasets that lead to inconsistent estimation of $\mathbf\beta_0$ using the proposed method in Section \ref{Doubly Robust Targeted-Federated Learning}.  To this end, we propose a bootstrap-based selection procedure that is designed to detect non-transportable source datasets, thereby numerically mitigating the risk of negative transfer. The extended framework consists of the following steps (Figure \ref{fig:Figure1}(a)):

\textbf{Step 1:} For the target dataset (i.e., $M = 1$), generate $B$ bootstrap samples $\mathcal{D}_1^{(b)}$ ($b = 1, \dots, B$). For each resampled dataset $\mathcal{D}_1^{(b)}$, fit the OR model $g_1^{(b)}(a, \mathbf{X}_i; \hat{\boldsymbol{\psi}}_1^{(b)})$ and the PS model $\pi_1^{(b)}(a, \mathbf{X}_i; \hat{\boldsymbol{\gamma}}_1^{(b)})$ and obtain the doubly robust estimator $\boldsymbol{\beta}_1^{(b)}$ by solving Equation (\ref{Eq:dr_est_target}). Meanwhile, calculate the covariate mean vector $\bar{\mathbf{X}}_1^{(b)}$ for each bootstrap sample. The collection of these mean vectors is then assembled into a mean matrix $\mathbf{R}_1 = [\bar{\mathbf{X}}_1^{(1)}, \dots, \bar{\mathbf{X}}_1^{(B)}]$, which is subsequently transferred to each source site $m \in \mathcal{S}$.

\textbf{Step 2:} For the source dataset $m \in \mathcal{S}$, generate bootstrap samples $\mathcal{D}_m^{(b)}$ ($b = 1, \dots, B$). For each bootstrap sample $\mathcal{D}_m^{(b)}$, estimate its OR model $g_m^{(b)}(a, \mathbf{X}_i; \hat{\boldsymbol{\psi}}_m^{(b)})$ and PS model $\pi_m^{(b)}(a, \mathbf{X}_i; \hat{\boldsymbol{\gamma}}_m^{(b)})$. Estimate the density ratio model $\rho_m^{(b)}(\mathbf{X}_i; \hat{\boldsymbol{\alpha}}_m^{(b)})$ parameterized by $\hat{\boldsymbol{\alpha}}_m^{(b)}$ using the mean matrix $\mathbf{R}_1$ from the target dataset. Finally, each source site transfers the full collection of augmentation terms $\{\mathbf{P}_m^{(b)}\}_{b=1}^{B}$ back to the target site, where they are used in the analysis of the target dataset.

\textbf{Step 3}: Set $w_m = 1$ and $w_{m'} = 0$ for all $m' \neq m, \; m\in\mathcal{K}$ in Equation (\ref{Eq:dr_targeted_federated}) to obtain $B$ bootstrap estimates $\boldsymbol{\beta}_m^{(b)}$ and compute the differences $\Delta_m^{(b)} = \boldsymbol{\beta}_m^{(b)} - \boldsymbol{\beta}_1^{(b)}$. A decision-making rule can be applied to determine the inclusion of the source site $m\in\mathcal{S}$ based on these differences. For instance, one may construct a Wald-type bootstrap 95\% confidence interval using bootstrap samples: if the 95\% CI covers zero, retain source dataset $m$ and specify $w_m$ based on the weighting strategy; otherwise, exclude source dataset $m$ and assign $w_m=0$.

A pseudo-algorithm is presented in Section 1.6 of the Supplementary Material. This learning procedure facilitates the exclusion of source datasets that produce estimates substantially deviating from the target-only estimate, thereby enhancing the robustness of the proposed learning process. Related oracle bootstrap-based variable selection theories have been established for LASSO-type regression \citep{bach2008bolasso,giurcanu2019bootstrapping}. Our proposed selection strategy may share a similar feature if a penalty is introduced to enforce similarity between $\boldsymbol{\beta}_m^{(b)}$ and $\boldsymbol{\beta}_1^{(b)}$. Conversely, if hypothesis testing is used within the proposed bootstrap procedure, bootstrap theory \citep{efron1994introduction} implies that non-transportable sources ($\boldsymbol{\beta}_m\neq\boldsymbol{\beta}_0$) would be detected with probability tending to one given regularity assumptions. We omit the formal theoretical assessment in this paper and instead focus primarily on demonstrating the practical utility of the proposed algorithm in targeted federated learning for studying HTEs. 

This framework offers several key advantages: i) it maintains communication efficiency (i.e., only one round of information exchange); ii) it is straightforward to implement and avoids the demanding need to transfer asymptotic influence functions; 
and iii) it demonstrates satisfactory numerical performance, as shown in Section \ref{SIMULATION}. The validity of this procedure relies on the consistency of the target-only estimator. Therefore, it is crucial to assess and validate the assumptions underlying the target-only estimate. We recommend using parametric models instead of machine learning methods for estimating the nuisance functions, as data sources in federated settings are often small and ML may perform poorly with limited samples \citep{chen2025effect}.

\section{SIMULATION STUDIES}
\label{SIMULATION}
\textbf{Transportable data sources.} We evaluated the finite-sample performance of the proposed targeted-federated estimator under the assumption that all source datasets were transportable, i.e., both assumptions and doubly robustness hold. We considered $K_s=10$ and $50$ external source datasets and examined the estimator performance under two target-sample sizes, i.e., $n_1=100, 300$. Corresponding source dataset sample sizes were drawn uniformly from $400$ to $600$ when $n_1=100$ and from $1500$ to $2000$ for $n_1=300$. Each dataset contained two continuous covariates, $X_1$ and $X_2$, and two binary covariates, $X_3$, and $X_4$. To emulate the real-world complexity of covariate distribution shifts across data sources, the target-site covariate distribution was fixed, whereas the source-site covariate means and Bernoulli probabilities were varied across source datasets. Potential outcomes and treatment assignment were generated from logistic models with prespecified coefficients. Full details of the covariate, treatment, and outcome data-generating mechanisms are provided in Section 2.1 of the Supplementary Material.

We considered three model specification settings. In Setting I, we studied the scenario where both the OR model and PS model were correctly specified. Setting II differed in that the PS model was misspecified by including $X_1$, $X_2$ only. Setting III differed from Setting I in that the OR model was misspecified by including $X_1$, $X_3$ only. For each setting, we examined estimation bias, Monte Carlo standard deviation (MCSD), the average bootstrap standard error (BSE), and the coverage probability of the $95\%$ CI based on $1000$ replicated Monte Carlo experiments. For all settings, we estimated the HTE of $X_1$ on the outcome $Y$.

We assessed and compared the performance of the following three estimators: i) a doubly robust estimator using the target dataset only; ii) a targeted-federated estimator that leverages data from all sites, assumes a constant density ratio of 1, without addressing the covariate distribution shift issue (Fed-SSnaive); iii) a targeted-federated estimator that leverages data from all sites, with a correctly specified density ratio model (Fed-SS). For Fed-based estimators, the source-specific weights were set to be proportional to their sample sizes (Section \ref{Doubly Robust Targeted-Federated Learning}).

\textbf{Non-transportable data sources. }We next evaluated the utility of the proposed bootstrap selection procedure in the presence of non-transportable source datasets. The covariate distributions and outcome generating mechanism followed the transportable sources scenario. We considered $n_1=400$ and $K_s=20$ source datasets, with source sample sizes uniformly drawn from $1500$ to $2000$. In this setting, the OR models for all datasets were misspecified by including $X_3$ and $X_4$ only, while the PS models were misspecified by including $X_3$ and $X_4$ only for source datasets with $n_m\geq Q_{0.5}$, i.e., those with sample sizes greater than the median among $\{n_m\}_{m=1}^{K_s}$. Consequently, half of the source datasets were non-transportable to the target data. In addition to the three candidate estimators above, we evaluated a targeted-federated estimator Fed-BS, which augments Fed-SS with the proposed bootstrap selection procedure.

\textbf{Results }We first evaluated the robustness of the estimators under three model specifications in scenarios with $K_s=10$ transportable data sources. In Settings I and II (Table \ref{tab:results_transportable}), all methods demonstrated satisfactory performance, exhibiting low estimation bias and satisfactory coverage. Among them, the target-only estimator showed slightly larger bias compared with the other two estimators under small sample size. However, under Setting III where the OR models were misspecified (Table \ref{tab:results_transportable}), the Fed-SSnaive estimator produced larger biases than both the target-only and Fed-SS estimators, indicating that Fed-SSnaive is more sensitive to covariate distribution shifts when the OR model is misspecified. Across settings, both Fed-SS and Fed-SSnaive estimators demonstrated substantially smaller MCSD than the target-only estimator, confirming the effectiveness of our proposed data integration methods in improving estimation efficiency. As the sample sizes increased, the BSE became closer to MCSD, and the Fed-SS estimator showed a desirable coverage probability, aligning well with the 95\% nominal level. We observed similar patterns with $K_s = 50$ (Section 2.2 of the Supplementary Material).

Table \ref{tab:results_nontransportable} summarizes the results in the presence of non-transportable data sources. All federated estimators showed substantially reduced estimation variability compared with the target-only estimator. However, both Fed-SS and Fed-SSnaive estimators, which incorporated non-transportable data sources, showed considerable estimation biases and coverage probabilities below the nominal $95\%$ level. In contrast, the Fed-BS estimator achieved markedly smaller bias and satisfactory coverage closer to the 95\% target. These findings underscore the importance of employing robust selection procedures in federated learning and support the validity of Fed-BS in identifying and excluding non-transportable data sources, thereby mitigating the risk of negative transfer.  

\textbf{Sensitivity analysis}. We conducted additional sensitivity analyses to assess the robustness of Fed-BS. First, we generated non-transportable source datasets using distinct potential-outcome models that violated Assumption 8 under mild, moderate, and large departures. Fed-BS substantially reduced bias, improved coverage, and lowered RMSE relative to the target-only estimator under moderate-to-large departures, whereas Fed-SS and Fed-SSnaive were sensitive to the inclusion of non-transportable sources. Under mild departures, the bootstrap procedure had reduced detection power and showed some undercoverage, although it remained more robust than the naive federated estimators. Second, we repeated the non-transportable-source simulation with $B=100$ bootstrap replicates and obtained results similar to those with $B=250$. We therefore recommend $B\in[150,300]$ as a practical range, with $B=250$ as the default. Detailed results and discussion of algorithmic stability, communication costs, and data privacy are provided in Section $2.3$ of the Supplementary Material.

\section{REAL DATA APPLICATION}
\label{REAL DATA APPLICATION}

Our research interest lies in analyzing recovery outcomes among older adults who experienced a hip fracture. Notably, hip fracture is widely recognized as one of the most debilitating injuries in older populations, often leading to a cascade of aging-related complications, such as disability, functional decline, loss of independence, reduced quality of life, and increased mortality \citep{beaupre2005recovery}. However, recovery trajectories vary considerably among patients, reflecting substantial heterogeneity in post-fracture outcomes \citep{falvey2023associations}. Therefore, it is important to identify potential driving factors that may partially explain poor recovery and help guide targeted interventions. 
 
\textbf{Study cohort, treatment, and outcome}. To maximize the utility of large-scale Medicare-linked data while acknowledging patient-level heterogeneity, our analysis specifically focuses on a minority subgroup: Black older adults who have been shown to have worse outcomes after hip fracture in general, but are largely unrepresented in major clinical studies making inference to more variable patient-centered outcomes difficult \citep{dy2016racial}.  We evaluated the potential differential impact of two common treatments for hip fracture: open reduction with internal fixation (OR/IF, $A=1$) vs hip arthroplasty (HA, $A=0$). Previous evidence suggests that HA has been associated with more favorable outcomes than internal fixation, particularly for postoperative functional scores and reoperation \citep{cui2020choice}. However, the substantial clinical heterogeneity among hip fracture survivors highlights the need to identify moderators that drive variation in surgical outcomes, motivating our investigation into the HTEs. The outcome was Days at Home (DAH), a patient-centered measure related to the concept of ``aging in place" that has been widely used in prior studies \citep{falvey2023associations, mutchie2023associations, shen2024analyzing}. 
DAH was defined as the number of days alive minus the total number of days spent in hospitals, skilled nursing facilities, hospital observation, and emergency department visits. To capture extremely unfavorable recovery, we defined the outcome $Y=1$ if total DAH was fewer than $10$ days or death occurred within 12 months after hip fracture, and $Y=0$ otherwise. The 2017 cohort, drawn from a 20\% national random sample of beneficiaries available to our team, served as the target dataset. Seven historical cohorts from 2010–2016, drawn from 5\% national random samples available to our collaborator, served as source datasets.

Eligible beneficiaries were aged 65 years or older and had a traumatic hip fracture. Hip fractures were identified using ICD-9 and ICD-10 diagnosis codes for femoral neck or intertrochanteric fractures, excluding fractures involving the pelvic rim or acetabulum. We excluded individuals who were not community-dwelling at baseline (i.e., long-stay nursing home residents prior to admission), those who died during their index admission for hip fracture, were discharged against medical advice, or were hospitalized on a U.S. territory (e.g., Puerto Rico). Additionally, we excluded individuals who did not maintain continuous Medicare fee-for-service enrollment for $6$ months prior to and $12$ months following the hip fracture admission. In addition, states with fewer than five eligible patients were also excluded to ensure data reliability and adequate representation. These criteria resulted in a final sample size of $1,830$ patients, with annual cohort sizes of $137$, $213$, $216$, $219$, $199$, $195$, $187$, and $464$ from $2010$ to $2017$, respectively. 

 \textbf{Covariates and potential effect modifiers}. To account for potential confounding effects, both the PS and OR models included a comprehensive set of factors, including demographic factors, hospital-level factors, and state of residence. Baseline patient-level factors included age group (0: ages 65–75; 1: 76–85; 2: 85+), sex (0: male; 1: female), pre-fracture diagnosis of Alzheimer's disease and related dementias (ADRD), Elixhauser comorbidity score during the 6 months before fracture, and total DAH during the 6 months before fracture. Hospital-level covariates included hospital bed-count tertile (0: $<230$ beds; 1: 231-424 beds; 2: $>425$ beds) and teaching hospital status. Sex and prior ADRD status were specified as potential effect modifiers because both have been strongly associated with post-fracture DAH outcomes \citep{mutchie2023associations}.

\begin{figure}
    \centering
    \includegraphics[width=1\linewidth]{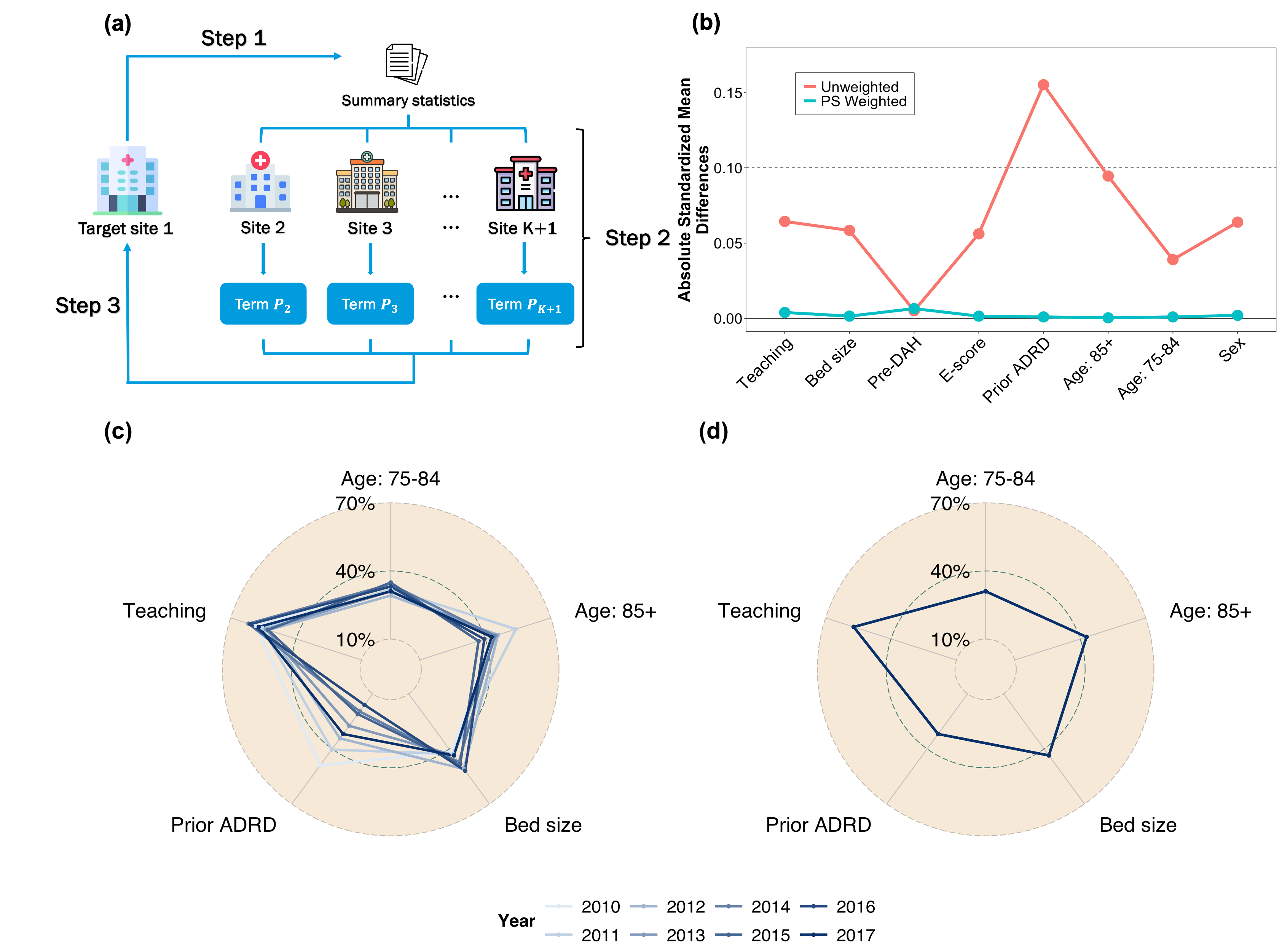}
    \caption{(a) Flowchart of the proposed targeted-federated framework. (b) Baseline covariate balance of the target dataset (year $2017$) was evaluated using absolute standardized mean differences before and after PS weighting. (c) \& (d) Radar plot of covariate distributional shifts from $2010$ to $2017$ before and after applying the density ratio weighting. Abbreviations: Teaching: teaching hospital status; Bed size: tertiles of hospital bed count (1: highest tertile; 0:otherwise); Pre-DAH: total DAH within 6 months prior to fracture; E-score: Elixhauser Comorbidity score; Prior ADRD: pre-fracture diagnosis of Alzheimer’s disease and related dementia (ADRD).}
    \label{fig:Figure1}
\end{figure}

\textbf{Model assessment}. To support the validity of applying our method, we conducted two assessments for assumptions. First, we examined the balance of baseline covariates in the target dataset. The love plot in Figure \ref{fig:Figure1}(b) revealed that PS weighting achieved good balance, with all absolute standardized mean differences below $0.1$. This supports the adequacy of the fitted PS model for the target-cohort analysis. Second, several covariates, including ADRD status and age category, showed distributional changes across calendar years, indicating covariate shifts between the target and source cohorts (Figure \ref{fig:Figure1}(c)). After density ratio weighting, these baseline covariates were well balanced across cohorts (Figure \ref{fig:Figure1}(d)), indicating that the density ratio model is likely to be correctly specified. Finally, the estimated propensity score distributions showed good overlap between treatment groups and no apparent practical violation of positivity (Figure 1 of the Supplementary Material). Further discussion of Assumptions 3, 4, 8 is provided in Section 3.1 of the Supplementary Material.

\begin{figure}
    \centering
    \includegraphics[width=0.75\linewidth]{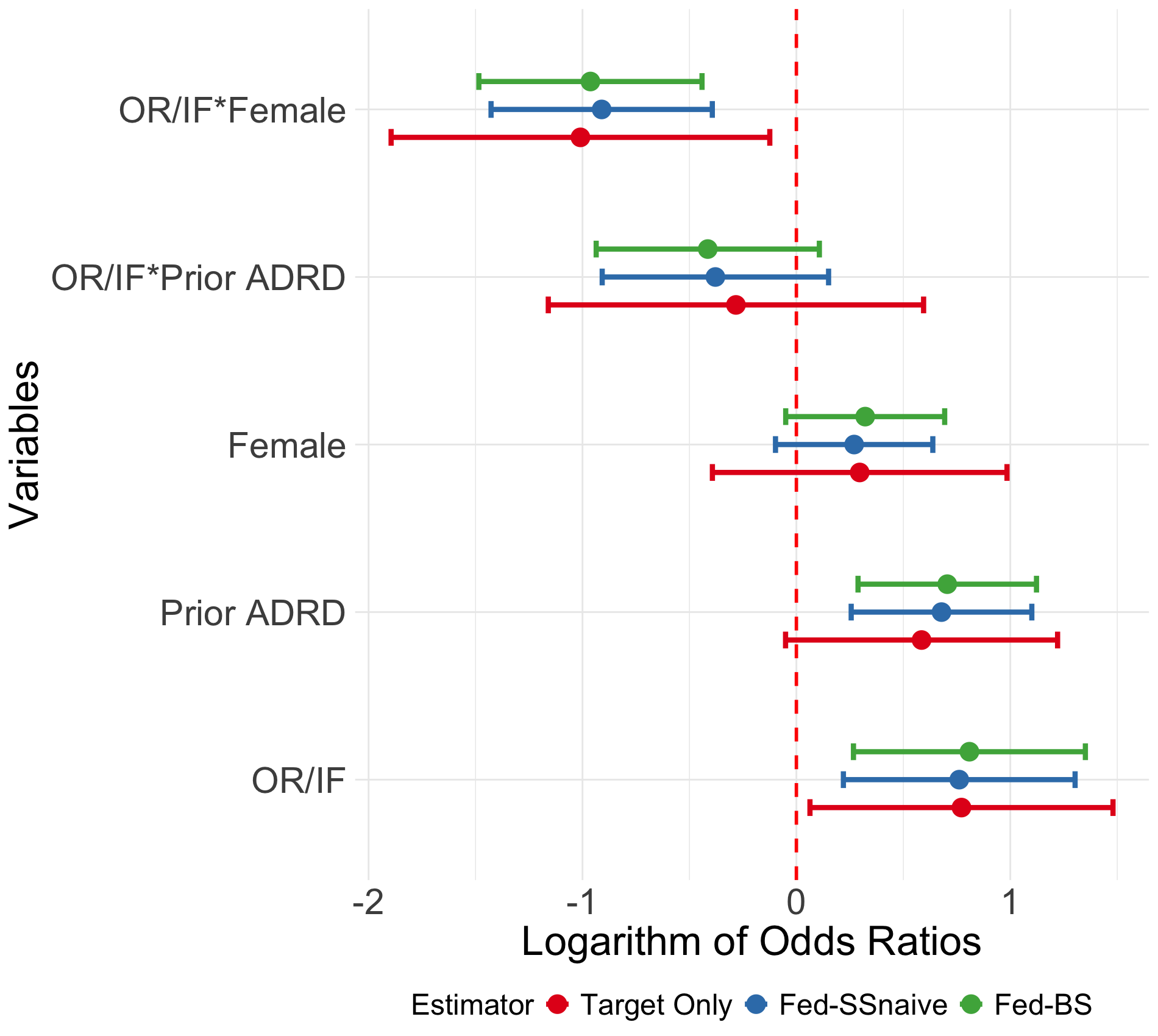}
    \caption{The estimated logarithm of odds ratios and corresponding 95\% bootstrap CIs for covariate effects in HTE models. The target dataset corresponds to the 2017 cohort. Using the proposed selection procedure,  all source datasets from 2010-2016 were identified as transportable datasets to the target cohort. Target-only: the non-federated estimator using the target dataset alone; Fed-SSnaive: the targeted-federated estimator without addressing covariate shift issue; Fed-BS: the proposed targeted-federated estimator enhanced by the bootstrap selection procedure.}
    \label{fig:forest_plot_main}
\end{figure}

\begin{figure}
    \centering
    \includegraphics[width=1\linewidth]{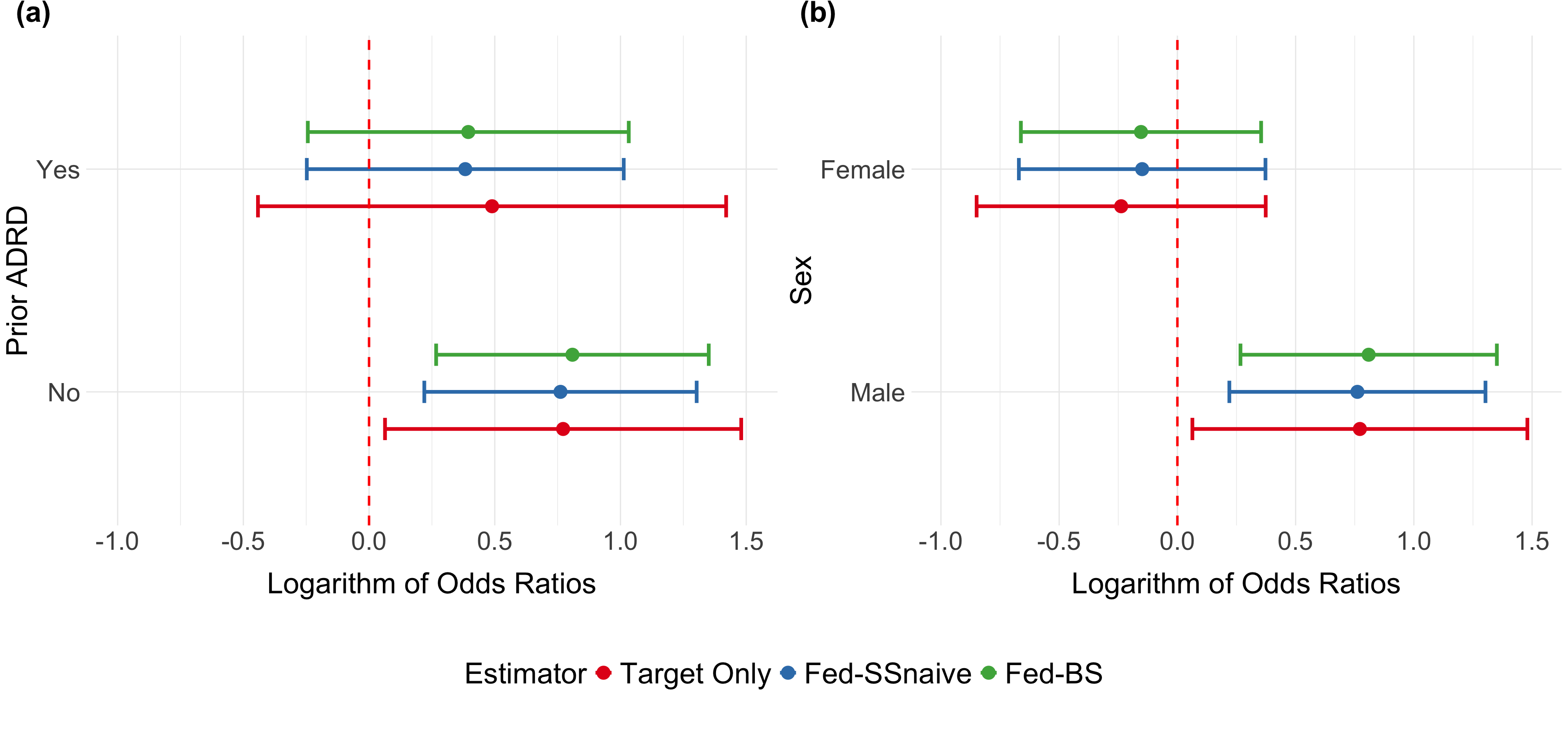}
    \caption{The estimated logarithm of odds ratios and corresponding 95\% bootstrap CIs for surgical effects (open reduction with internal fixation vs. hip arthroplasty) among different subgroups: (a) ADRD or not; (b) Female and male. Target-only: the non-federated estimator using the target dataset alone; Fed-SSnaive: the targeted-federated estimator without addressing covariate shift issue; Fed-BS: the proposed targeted-federated estimator enhanced by the bootstrap selection procedure.}
    \label{fig:forest_plot_subgroup}
\end{figure}

\textbf{Results}. In the targeted-federated analysis, the proposed bootstrap procedure described in Section \ref{method extension} identified all $2010-2016$ cohorts as transportable to the $2017$ target cohort. The estimated log odds ratios from three approaches, i.e., Target only, Fed-SSnaive, and Fed-BS, are summarized in Figure \ref{fig:forest_plot_main}, with corresponding 95\% confidence intervals constructed via bootstrapping. All three estimators detected a significant positive main effect from OR/IF and a significant negative interaction effect from OR/IF by sex. These findings suggest that OR/IF is more likely to result in unfavorable recovery than HA for Black patients without pre-fracture diagnosis of ADRD, and that sex acts as a significant effect modifier, leading to differential treatment effects (Figure \ref{fig:forest_plot_subgroup}). Additionally, both the Fed-BS and Fed-SSnaive approaches captured a significant association between pre-fracture ADRD and worse recovery outcome, whereas the target-only estimator did not identify this effect as significant. In general, Fed-BS and Fed-SSnaive produced narrower confidence intervals, indicating the utility of federated learning in improving estimation efficiency. However, in this particular case study, we did not observe a significant difference in point estimates or uncertainty between the Fed-BS and Fed-SSnaive estimators. Moreover, we observed consistent results in sensitivity analyses where we considered fewer than $5$ days or $20$ days as thresholds in the definition of the outcome, in place of the $10$-day cutoff used in the primary analysis (Section 3.2 of the Supplementary Material).

\section{DISCUSSION}
\label{DISCUSSION}
In this paper, we developed a novel targeted-federated learning framework for estimating HTEs. This method addresses an important methodological gap in the federated learning literature, where existing approaches have limited their scope to prediction, classification, or ATE-based causal inference. Our framework requires only a single round of communication across data sources and achieves double robustness while mitigating negative data transfer, making it robust and communication-efficient for real-world federated applications. Findings from our data application suggest that HA may result in significantly better post-fracture recovery than OR/IF procedure. In addition, pre-fracture ADRD status appears to be a potential prognostic factor contributing to differential surgery effects. These findings offer evidence to improve our understanding of the heterogeneous impact of surgical treatments for older adults with a hip fracture. 

Despite these advancements, the current framework has several limitations that warrant further investigation. One limitation is the potential for unmeasured confounding, particularly because surgery type is often dictated by fracture characteristics (e.g., severity), which are not consistently captured in Medicare claims data. Additionally, a formal theoretical assessment of the bootstrap-based selection algorithm in targeted-federated learning remains to be developed. In practice, the choice of $B$ reflects a balance among statistical stability, communication volume, and privacy considerations, as larger $B$ may increase the risk of privacy leakage. Future work can focus on mitigating data privacy risks. In addition, diagnosis of the working structural model, as well as nonparametric estimation of treatment effects and the variance of treatment effects \citep{levy2021fundamental}, remain important directions for future research in federated learning settings.

\bigskip

\spacingset{1.45}
\bibliographystyle{plain} 
\bibliography{biomsample_bib}

\begin{table}[H]
\centering
\caption{Evaluation of estimators under the setting with transportable source data. $K_s=10$}
\label{tab:results_transportable}
\resizebox{\textwidth}{!}{%
\begin{tabular}{@{}lcccccccccccccccccccc@{}}
\toprule
 & \multicolumn{5}{c}{Intercept} & \multicolumn{5}{c}{$A_i$} & \multicolumn{5}{c}{$X_{1i}$} & \multicolumn{5}{c}{$A_i*X_{1i}$} \\ \cmidrule(l){2-21}
 & Bias & MCSD & BSE & RMSE & CP & Bias & MCSD & BSE & RMSE & CP & Bias & MCSD & BSE & RMSE & CP & Bias & MCSD & BSE & RMSE & CP \\ \midrule
$n_1=100$ & & & & & & & & & & & & & & & & & & & & \\ \cmidrule(r){1-1}
Setting I & & & & & & & & & & & & & & & & & & & & \\
Target-only & -1.8 & 37.4 & 41.2 & 37.4 & 95.8 & 1.7 & 51.7 & 59.0 & 51.7 & 96.3 & 2.4 & 47.7 & 52.1 & 47.8 & 97.4 & 4.8 & 64.6 & 75.2 & 64.8 & 96.9 \\
Fed-SS & -1.7 & 34.4 & 35.8 & 34.4 & 95.6 & 1.6 & 47.2 & 51.3 & 47.2 & 96.1 & 1.8 & 35.3 & 38.6 & 35.3 & 97.2 & 2.9 & 49.7 & 56.9 & 49.8 & 97.1 \\
Fed-SSnaive & -1.6 & 34.5 & 35.9 & 34.5 & 95.7 & 1.6 & 47.2 & 51.3 & 47.2 & 96.0 & 1.9 & 35.6 & 39.2 & 35.7 & 97.5 & 2.6 & 49.7 & 56.9 & 49.8 & 97.1 \\
Setting II & & & & & & & & & & & & & & & & & & & & \\
Target-only & -2.2 & 36.6 & 39.4 & 36.7 & 95.8 & 2.2 & 51.6 & 56.6 & 51.6 & 96.1 & 1.8 & 49.1 & 49.1 & 49.1 & 97.3 & 4.6 & 66.3 & 71.1 & 66.5 & 97.1 \\
Fed-SS & -1.8 & 34.4 & 35.8 & 34.4 & 95.8 & 1.8 & 47.4 & 51.3 & 47.4 & 96.0 & 1.8 & 35.3 & 38.7 & 35.3 & 97.1 & 3.0 & 49.7 & 56.8 & 49.8 & 97.1 \\
Fed-SSnaive & -1.8 & 34.5 & 35.9 & 34.5 & 95.9 & 1.9 & 47.3 & 51.4 & 47.3 & 96.0 & 1.9 & 35.6 & 39.1 & 35.7 & 97.4 & 2.7 & 49.7 & 56.8 & 49.8 & 97.0 \\
Setting III & & & & & & & & & & & & & & & & & & & & \\
Target-only & -2.4 & 38.3 & 42.4 & 38.4 & 96.6 & 2.5 & 53.3 & 62.3 & 53.4 & 97.2 & 1.5 & 52.6 & 54.1 & 52.6 & 97.5 & 4.2 & 72.1 & 82.6 & 72.2 & 96.1 \\
Fed-SS & -1.9 & 32.4 & 33.5 & 32.5 & 96.5 & 1.1 & 45.2 & 49.6 & 45.2 & 96.6 & 1.8 & 34.9 & 38.0 & 34.9 & 96.6 & 0.5 & 51.1 & 58.9 & 51.1 & 96.6 \\
Fed-SSnaive & -1.4 & 32.5 & 33.7 & 32.5 & 96.5 & 5.6 & 45.0 & 49.0 & 45.3 & 94.8 & 3.9 & 35.4 & 38.7 & 35.6 & 95.7 & -7.1 & 50.2 & 57.1 & 50.7 & 95.1 \\ \midrule
$n_1=300$ & & & & & & & & & & & & & & & & & & & & \\ \cmidrule(r){1-1}
Setting I & & & & & & & & & & & & & & & & & & & & \\
Target-only & -1.6 & 19.2 & 20.0 & 19.3 & 94.6 & 1.2 & 27.0 & 28.2 & 27.0 & 96.2 & 0.5 & 22.9 & 24.5 & 22.9 & 95.8 & 1.5 & 32.1 & 34.0 & 32.1 & 95.4 \\
Fed-SS & -1.3 & 18.0 & 18.6 & 18.0 & 95.0 & 0.6 & 25.2 & 26.3 & 25.2 & 95.9 & 1.1 & 17.4 & 18.7 & 17.4 & 95.9 & 1.0 & 25.9 & 27.1 & 25.9 & 96.5 \\
Fed-SSnaive & -1.3 & 18.0 & 18.6 & 18.0 & 95.0 & 0.6 & 25.2 & 26.3 & 25.2 & 95.8 & 1.2 & 17.6 & 18.8 & 17.6 & 95.7 & 0.9 & 25.9 & 27.1 & 25.9 & 96.2 \\
Setting II & & & & & & & & & & & & & & & & & & & & \\
Target-only & -1.6 & 18.9 & 19.7 & 19.0 & 95.5 & 1.0 & 27.6 & 27.8 & 27.6 & 96.1 & 0.4 & 22.8 & 23.6 & 22.8 & 95.8 & 1.9 & 31.9 & 33.0 & 32.0 & 95.4 \\
Fed-SS & -1.3 & 18.0 & 18.6 & 18.0 & 95.1 & 0.6 & 25.2 & 26.3 & 25.2 & 95.8 & 1.1 & 17.4 & 18.7 & 17.4 & 95.9 & 1.0 & 25.9 & 27.1 & 25.9 & 96.6 \\
Fed-SSnaive & -1.3 & 18.0 & 18.6 & 18.0 & 95.0 & 0.6 & 25.2 & 26.3 & 25.2 & 95.8 & 1.1 & 17.6 & 18.8 & 17.6 & 96.0 & 0.9 & 25.9 & 27.0 & 25.9 & 96.3 \\
Setting III & & & & & & & & & & & & & & & & & & & & \\
Target-only & -1.6 & 19.1 & 20.1 & 19.2 & 94.9 & 1.4 & 27.3 & 28.8 & 27.3 & 96.0 & 0.4 & 23.0 & 24.9 & 23.0 & 95.7 & 1.7 & 34.3 & 36.4 & 34.3 & 95.6 \\
Fed-SS & -1.3 & 17.2 & 17.6 & 17.2 & 94.4 & 0.7 & 24.6 & 25.5 & 24.6 & 95.5 & 1.1 & 17.3 & 18.5 & 17.3 & 95.9 & 1.3 & 26.3 & 27.7 & 26.3 & 96.4 \\
Fed-SSnaive & -0.9 & 17.2 & 17.6 & 17.2 & 94.5 & 5.7 & 24.4 & 25.3 & 25.1 & 94.0 & 2.7 & 17.5 & 18.7 & 17.7 & 95.7 & -6.7 & 25.8 & 27.3 & 26.7 & 94.5 \\ \bottomrule
\end{tabular}%
}
\footnotesize This table presents the performance of three estimators: Target-only: the non-federated estimator using the target dataset alone; Fed-SS: the proposed targeted-federated estimator; Fed-SSnaive: the targeted-federated estimator without addressing the covariate shift issue. MCSD: Monte Carlo standard deviation; BSE: bootstrap standard error; RMSE: root mean squared error; CP: 95\% coverage probability. All values were multiplied by $100$.
\end{table}

\begin{table}[H]
\centering
\caption{Evaluation of estimators in the presence of non-transportable source data. $B=250$}
\label{tab:results_nontransportable}
\resizebox{\textwidth}{!}{%
\begin{tabular}{@{}lcccccccccccccccccccc@{}}
\toprule
 & \multicolumn{5}{c}{Intercept} & \multicolumn{5}{c}{$A_i$} & \multicolumn{5}{c}{$X_{1i}$} & \multicolumn{5}{c}{$A_i*X_{1i}$} \\ \cmidrule(l){2-21}
 & Bias & MCSD & BSE & RMSE & CP & Bias & MCSD & BSE & RMSE & CP & Bias & MCSD & BSE & RMSE & CP & Bias & MCSD & BSE & RMSE & CP \\ \cmidrule(l){2-21}
Target-only & 0.0 & 16.4 & 16.5 & 16.4 & 95.3 & -1.0 & 24.5 & 24.9 & 24.5 & 95.4 & 0.0 & 18.7 & 18.7 & 18.7 & 95.9 & 3.2 & 30.6 & 32.2 & 30.8 & 97.1 \\
Fed-SS & -2.9 & 14.4 & 14.4 & 14.7 & 95.1 & 18.9 & 20.6 & 20.8 & 28.0 & 87.3 & 0.5 & 3.3 & 3.6 & 3.3 & 97.1 & -8.9 & 5.5 & 5.9 & 10.5 & 65.7 \\
Fed-SSnaive & -2.4 & 14.3 & 14.4 & 14.5 & 95.3 & 21.1 & 20.5 & 20.7 & 29.4 & 85.1 & 0.4 & 2.8 & 3.1 & 2.8 & 97.3 & -12.0 & 4.2 & 4.7 & 12.7 & 31.5 \\
Fed-BS & -0.8 & 14.4 & 14.4 & 14.4 & 95.7 & 3.4 & 21.2 & 20.9 & 21.5 & 93.8 & -0.2 & 3.9 & 3.8 & 3.9 & 95.4 & -1.0 & 8.7 & 7.8 & 8.8 & 92.2 \\ \bottomrule
\end{tabular}%
}
\footnotesize This table presents the performance of four estimators: Target-only: the non-federated estimator using the target dataset alone; Fed-SS: the proposed targeted-federated estimator; Fed-SSnaive: the targeted-federated estimator without addressing the covariate shift issue; Fed-BS: the proposed targeted-federated estimator enhanced by the bootstrap selection procedure. MCSD: Monte Carlo standard deviation; BSE: bootstrap standard error; RMSE: root mean squared error; CP: 95\% coverage probability. All values were multiplied by $100$.
\end{table}

\end{document}